\documentclass[conference]{IEEEtran}
\IEEEoverridecommandlockouts
\usepackage{cite}
\usepackage{amsmath,amssymb,amsfonts}
\usepackage{algorithmic}
\usepackage{graphicx}
\usepackage{textcomp}
\usepackage{xcolor}
\def\BibTeX{{\rm B\kern-.05em{\sc i\kern-.025em b}\kern-.08em
    T\kern-.1667em\lower.7ex\hbox{E}\kern-.125emX}}

\begin{document}

\title{Enhancing Code Consistency in AI Research with Large Language Models and Retrieval-Augmented Generation}


\author{\IEEEauthorblockN{Rajat Keshri, Arun George Zachariah, Michael Boone \\
\textit{NVIDIA} \\
rkeshri@nvidia.com, azachariah@nvidia.com, mboone@nvidia.com
}
}

\maketitle

\begin{abstract}
Ensuring that code accurately reflects the algorithms and methods described in research papers is critical for maintaining credibility and fostering trust in AI research. This paper presents a novel system designed to verify code implementations against the algorithms and methodologies outlined in corresponding research papers. Our system employs Retrieval-Augmented Generation to extract relevant details from both the research papers and code bases, followed by a structured comparison using Large Language Models. This approach improves the accuracy and comprehensiveness of code implementation verification while contributing to the transparency, explainability, and reproducibility of AI research. By automating the verification process, our system reduces manual effort, enhances research credibility, and ultimately advances the state of the art in code verification.
\end{abstract}


\section{Introduction}
Reproducibility is a critical challenge in artificial intelligence (AI) and machine learning (ML) research, essential for ensuring the reliability and trustworthiness of published findings \cite{Semmelrock_2024} \cite{Hutson_2018}. Inconsistent replication of results across implementations or codebases can hinder scientific progress and undermine confidence in research outcomes \cite{Pineau_2019}. Studies reveal that a significant portion of published AI research suffers from reproducibility issues, with substantial discrepancies between the methodology described in papers and the codebases provided \cite{Gundersen_2018}. As experiments and data processing in ML become more complex, even minor discrepancies in code or parameter settings can lead to substantial deviations in results, making verification both difficult and time-consuming \cite{Tatman_2018}.

Traditional reproducibility efforts rely on manual verification, where researchers cross-check code against described methodologies \cite{Rider_2017} \cite{Raff_2019}. However, this process is labor-intensive, subjective, and prone to error, highlighting the need for automated systems that can systematically assess alignment between research papers and codebases. Recent advances in Large Language Models (LLMs) show promise in this domain, as they can analyze both natural language and programming code, making them suitable for examining consistency across research artifacts \cite{Lu_2023}.

This paper introduces an automated system that combines LLMs with Retrieval-Augmented Generation (RAG) \cite{Lewis_2020} to verify alignment between research papers and code implementations. By automatically extracting key research elements, such as model architecture, hyperparameters, and algorithms, and comparing these with actual code, our system reduces manual effort and minimizes the risk of human error. RAG enhances the LLM’s retrieval accuracy, addressing issues like hallucination, where models generate plausible but incorrect responses.

Through this automated verification process, we aim to provide researchers and reviewers with a reliable tool for validating code consistency, thereby promoting research transparency and integrity. Related works use language models to generate and maintain data cards~\cite{Truong_DataCards_2026} and model cards~\cite{Keshri_ModelCards_2026}, complementing our focus on paper-code consistency. Scalable cloud execution of machine learning tasks~\cite{Masson_Cloud_2026} offers a potential deployment approach for this verification workflow.

\section{System Design}
\begin{figure}[h]
    \centering
    \includegraphics[width=0.5\textwidth]{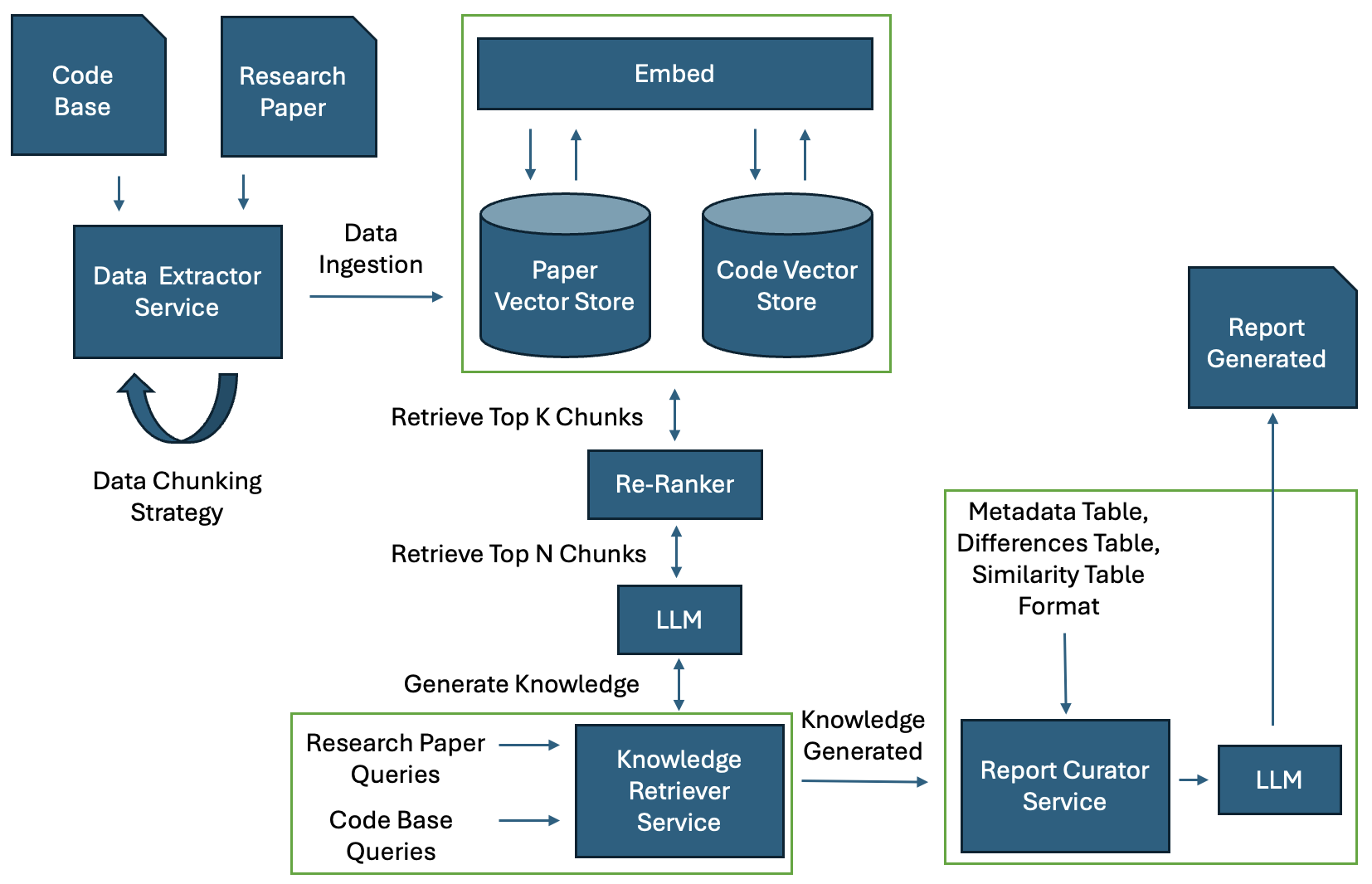}
    \caption{System Architecture}
    \label{System_Architecture}
\end{figure}

\subsection{Architecture}
Our proposed system architecture, shown in Fig.~\ref{System_Architecture}, comprises of four main components: \textit{Data Extractor Service}, \textit{Embed Module (Paper Vector Store and Code Vector Store)}, \textit{Knowledge Retriever Service}, and \textit{Report Curator Service}. Together, these components form a streamlined pipeline for parsing, extracting, storing, retrieving, and verifying research content, ultimately producing a report that highlights any discrepancies between the paper and the codebase.

\noindent \textbf{1. Data Extractor Service:} 
The Data Extractor Service is the initial module in the system pipeline, tasked with processing inputs from users, which include a research paper in PDF format and code files as a compressed ZIP archive. This service converts the research paper from PDF to markdown format, enabling efficient segmentation and processing of text. The markdown format allows easy access to sections, paragraphs, and headings, which are subsequently segmented into coherent chunks based on the structure and content flow of the document. The paper is divided according to predefined headings such as \textit{“Introduction”}, \textit{“Methodology”}, \textit{“Results”}, and \textit{“Discussion}”, which helps to ensure that extracted content remains logically organized and relevant to specific areas of the paper.

Similarly, the codebase is processed by the Data Extractor Service to enable granular analysis. The service decompresses the ZIP archive and splits the code files into chunks on a file-by-file basis. Each code file is individually parsed, and larger files are segmented into smaller, manageable sections based on code structure, such as function or class definitions. By isolating these components, the service captures relevant elements of the codebase, including method definitions, parameter settings, and configuration files. The chunked and segmented data, from both the paper and the codebase, is then prepared for vectorization and storage in subsequent stages of the pipeline. The Data Extractor Service thus lays the foundation for a systematic, component-wise examination of the research and implementation, facilitating a thorough verification process downstream.

\noindent \textbf{2. Embed Module - Paper Vector Store and Code Vector Store:}
The Embed Module is critical to transforming the segmented data into machine-readable representations, stored in two distinct vector repositories: the Paper Vector Store and the Code Vector Store. These vector stores serve as foundational databases, enabling efficient semantic similarity searches between the research paper and the codebase.
    
In the Paper Vector Store, the vector representations are derived from the research paper’s chunks. The system applies embedding techniques from advanced Natural Language Processing (NLP) models to create high-dimensional vectors that capture the semantic meaning of text chunks. By using large language model (LLM) embeddings, the Paper Vector Store achieves a comprehensive representation of the paper’s content, encapsulating key details such as theoretical background, methodologies, and experimental setups. These embeddings are stored with metadata that links each vector back to its originating section within the paper, allowing for easy retrieval and contextual alignment with corresponding code sections.
    
The Code Vector Store operates in parallel with the Paper Vector Store, focusing on vectorized representations of code chunks. The Code Vector Store vectorizes various elements of the codebase, including functions, classes, and configuration settings, converting them into embeddings that capture the structure, functionality, and purpose of each code segment. These vector representations are crucial for identifying similarities between code implementation details and the descriptions within the research paper. Additionally, by organizing the code in a structured format, the system supports efficient data retrieval, enabling quick and precise searches within the codebase.
    
The Embed Module thus enables efficient cross-referencing between the research paper and codebase, serving as the foundation for downstream similarity matching. The dual storage of vectorized representations in the Paper Vector Store and Code Vector Store facilitates alignment and discrepancy analysis, enhancing the system’s ability to verify that code implementations accurately reflect research claims.

\noindent \textbf{3. Knowledge Retriever Service:}
The Knowledge Retriever Service leverages Retrieval-Augmented Generation (RAG) techniques to gather and synthesize relevant information from both the research paper and the codebase, focusing on predefined queries tailored to extract specific aspects of the research methodology and implementation. This service serves as an intermediary layer that enhances the alignment analysis by addressing targeted questions that are central to verifying research reproducibility. Examples of these queries include, “What is the model architecture described in the paper?”, “What hyperparameters are suggested for training?”, and “What training algorithm is used?” These queries are designed to extract key elements from both the paper and the code, guiding the system toward essential alignment factors such as model specifications, data preprocessing steps, and experimental conditions.
    
Upon processing a query, the Knowledge Retriever Service accesses both the Paper and Code Vector Stores, retrieving the top K relevant chunks for each query. These initial results are re-ranked based on their relevance and similarity to the query, using advanced re-ranking algorithms to ensure that the most contextually accurate information is prioritized. Once re-ranked, the top results are inputted into the LLM, which synthesizes the information and generates a comprehensive summary for each query, covering both the research paper and codebase perspectives. By retrieving and synthesizing information for targeted queries, the Knowledge Retriever Service enhances the system’s accuracy in identifying alignment or discrepancies, enabling a more detailed and focused assessment of consistency.
    
The Knowledge Retriever Service thus enriches the system’s ability to perform precise alignment checks, allowing it to verify whether critical aspects of the research paper are reflected in the codebase. This targeted retrieval and synthesis process serves as an additional layer of validation, addressing specific research elements in a structured manner and ensuring that the verification process covers all significant aspects of the research.

\noindent \textbf{4. Report Curator Service:}
The Report Curator Service is responsible for compiling the final output of the system, presenting a structured and comprehensive report that summarizes the consistency analysis findings. This service is designed to organize the extracted and synthesized information into a user-friendly report format, allowing researchers, reviewers, and other stakeholders to quickly understand the degree of alignment between the research paper and the codebase.
    
Using predefined templates, the Report Curator Service organizes the report into key sections, including metadata, alignment analysis, and discrepancy summaries. The metadata section provides basic information about the research paper and codebase, including titles, authors, publication dates, and any other relevant identifiers. The alignment analysis section includes tables and similarity scores that indicate the degree of consistency between specific sections of the paper and corresponding code elements. These tables are accompanied by a difference analysis that highlights areas where discrepancies were found, such as mismatches in hyperparameters, data handling techniques, or model architecture.
    
In addition to discrepancy tables, the report may include similarity tables for sections that demonstrate strong alignment, offering readers a clear view of sections where the research claims are well-supported by the code. This structured format provides a balanced view, showcasing both strengths and weaknesses in the code-paper alignment. Furthermore, the system’s alignment check helps assess the overall reliability and reproducibility of the research, offering a quantified view of consistency through an alignment score. This score reflects the extent to which the codebase adheres to the research claims, providing a useful indicator for authors and reviewers.
    
The Report Curator Service thus adds a final layer of structure to the verification process, creating a clear and informative report that highlights the degree of alignment and reproducibility within the research. By presenting findings in an accessible and organized manner, the Report Curator Service promotes transparency and accountability, encouraging researchers to provide code that accurately reflects their research methodologies and results.

\subsection{Workflow}
The overall system workflow begins with user-provided inputs (research paper in PDF format and codebase in ZIP format), processed initially by the Data Extractor Service. After data extraction and chunking, the Embed Module vectorizes and stores the processed content in the Paper and Code Vector Stores. The Knowledge Retriever Service then executes targeted queries to extract and synthesize information from these vector stores, comparing research claims with code implementations. Finally, the Report Curator Service compiles the synthesized information into a structured report, offering a clear summary of alignment findings.

\subsection{Implementation}
The system is implemented using Streamlit for the frontend, providing an intuitive user interface where researchers can upload files and view results. On the backend, llama-index \cite{Liu_2022} is employed for indexing and retrieval, while NVIDIA’s NeMo toolkit \cite{NVIDIA_2019} integrates the Llama-3.1-8b-instruct \cite{Meta_2024} model to handle queries and generate comparisons.

Embedding generation is handled using NV-Embed-QA \cite{Lee_2024}, with vector representations stored in the Chroma Vector Database. For the retrieval of relevant chunks, the NVIDIA Re-rank \cite{NVIDIA_2024} module refines results to ensure accurate comparisons between paper sections and code implementations.

The final verification report is presented in a clear and structured format, summarizing how well the codebase implements the algorithms described in the paper and highlighting any discrepancies.

\section{Demonstration Scenarios}
\begin{figure}[h]
    \centering
    \includegraphics[width=0.46\textwidth]{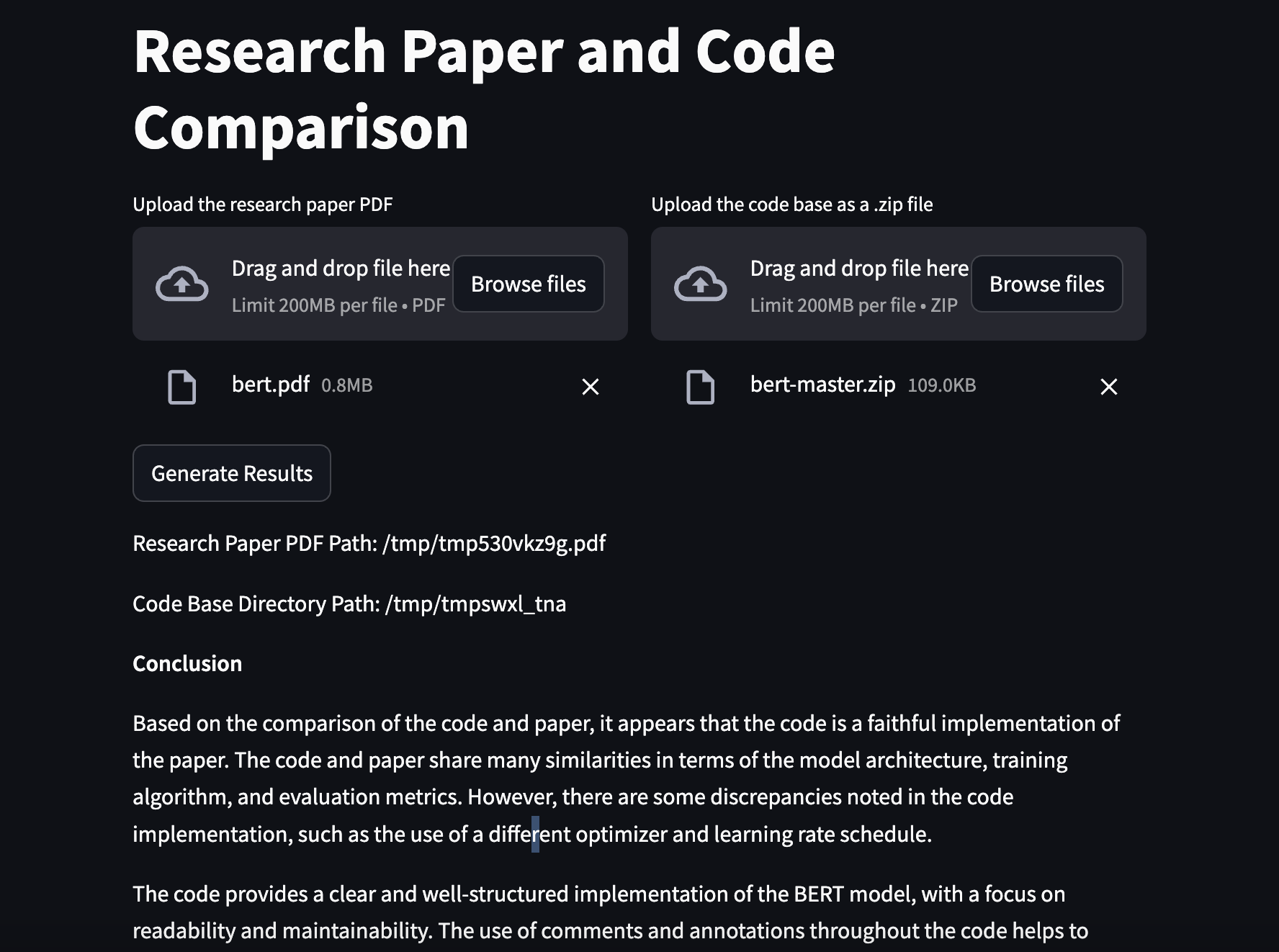}
    \caption{System UI}
    \label{System_UI}
\end{figure}
To illustrate the functionality of our system, we present two primary usage scenarios: \textit{Off-The-Shelf Model Verification} and \textit{Custom Model Verification}. These scenarios demonstrate how users can interact with the system to ensure that a research paper’s claims are accurately implemented in its corresponding codebase. By following these scenarios, users can verify the alignment between the paper and the codebase, helping to ensure the reproducibility and integrity of research findings.

\noindent \textbf{1. Off-The-Shelf Model Verification:} 
In this scenario, the user begins by uploading a research paper that describes a well-known model, such as BERT \cite{Kenton_2019}, along with its corresponding codebase. This scenario is particularly relevant for models that are widely used and have established standards, making them suitable for direct verification against expected implementations. Once the user has uploaded the research paper (in PDF format) and code files (in ZIP format), the system initiates its data extraction and vectorization process. The Data Extractor Service parses the paper, identifying key sections like \textit{“Methodology”} and \textit{“Results”}, and segments the content accordingly. Similarly, the codebase is decompressed and broken down into file-based chunks, isolating critical components such as model definitions, hyperparameters, and configuration files.

As the system processes both inputs, the Knowledge Retriever Service extracts specific details from both the paper and codebase that are relevant to the BERT model’s standard attributes. For instance, it retrieves data on model architecture, tokenization methods, layers, hyperparameters, and optimization settings, which are then re-ranked based on their relevance to the initial queries. The system subsequently generates a report comparing these extracted attributes, highlighting any inconsistencies found between the paper’s claims and the actual implementation in the code. The generated report provides a structured overview of key findings, showing matched elements, such as identical hyperparameter values, as well as detected differences, such as a mismatch in model depth or activation functions.

Through the interactive user interface, users can view these findings in a structured format. For example, sections within the report might show side-by-side comparisons of expected vs. actual hyperparameters, training epochs, or model configurations. The user can navigate the report, explore individual discrepancies, and dive into areas of interest by clicking on specific sections. This scenario allows users to conduct a high-level inspection of the implementation’s alignment with the paper, ensuring that the code meets expected standards for reproducibility.

\noindent \textbf{2. Custom Model Verification:} In this scenario, the user uploads a research paper and its associated codebase for a unique model or experimental setup that may not adhere to standardized configurations. Unlike Off-The-Shelf Model Verification, which focuses on pre-established models, Custom Model Verification emphasizes validating novel approaches and experimental details that are unique to the user’s research. After the user provides the research paper and codebase files, the system performs the same parsing and segmentation process, isolating key sections in both the text and code.

The Knowledge Retriever Service then applies targeted queries designed to extract specific aspects of custom implementations, such as unique algorithms, data preprocessing steps, training strategies, or custom evaluation metrics. The system gathers these details, focusing on aspects that are essential to reproducing the custom model’s results. For instance, if the paper introduces a novel training algorithm, the system compares the implementation in the code with the algorithm’s description in the paper, identifying any discrepancies in the steps, parameters, or methods used.

The generated report offers a detailed analysis, structured to help users easily identify critical points of alignment or deviation. For instance, the report may contain tables or highlighted text that contrasts the expected implementation details with the actual code, bringing attention to any inconsistencies or missing components. The user interface enables users to interact with these results intuitively, viewing detailed explanations for each discrepancy and exploring sections dedicated to methodology, algorithm implementation, or training configurations. By providing this detailed comparison, the system assists users in verifying the custom model’s reproducibility, which is often crucial for experimental research.

In both scenarios, users interact with the system through an intuitive, web-based interface, seen in Fig.~\ref{System_UI}. They begin by uploading the required files after which the system processes and analyzes the content automatically. Once the verification is complete, the system presents the results in an organized report format accessible through the same interface. The interactive interface allows users to navigate through various sections of the report, zooming in on specific elements such as model architecture, hyperparameters, training steps, or evaluation metrics.

The user interface is designed to support easy exploration of the results, with collapsible sections and clickable links that allow users to delve deeper into specific findings. Each section of the report provides a detailed view of the analysis, with tables, charts, and annotations that highlight areas of consistency and inconsistency. This structured presentation ensures that users can quickly identify and address any issues in the code-paper alignment, streamlining the verification process.

\section{Conclusion}
This paper introduces an automated system using LLMs and RAG to verify consistency between research papers and their codebases. By automating extraction and comparison, it addresses the reproducibility challenge in AI and data engineering research. The system generates detailed reports that highlight discrepancies, reduces manual effort, and enhances transparency in academic publishing. Our demonstration will showcase how this tool simplifies verification, providing researchers, reviewers, and conference organizers with a practical means to ensure research reliability.

\end{document}